\documentclass{rspublic}

\begin{document}
\title{Cosmic Superstrings}
\author{Mairi Sakellariadou}
\affiliation{Department of Physics, King's College London\\
University of London, Strand, WC2R 2LS, UK}

\label{firstpage}

\maketitle

\begin{abstract}{Cosmic superstrings, F/D strings, $(p,q)$-bound states, 
brane inflation}

Cosmic superstrings are expected to be formed at the end of brane
inflation, within the context of brane-world cosmological models
inspired from string theory. By studying properties of cosmic
suprestring networks, and comparing their phenomenological
consequences against observational data, we aim at pinning down the
successful and natural inflationary model and get an insight into the
stringy description of our universe.

\end{abstract}

\section{Introduction}

Inflation was proposed (Guth 1981; Linde 1982) in the eighties as a
simple and elegant solution to the shortcomings of the hot big bang
model. Inflation provides a solution to the flatness and horizon
problems within the framework of quantum field theory and general
relativity, while it dilutes any undesired relics (e.g., monopoles)
from possible extensions of the standard model.  Inflation essentially
consists of a phase of accelerated expansion which took place at a
very high energy scale.  Thus, by construction, inflation can liberate
the standard model of cosmology (i.e., the hot big bang model) from
the requirement of special initial conditions. In addition,
amplification of the quantum fluctuations of the inflaton field, which
drives inflation, offers a simple mechanism for the origin of the
initial density fluctuations, which via gravitational instability
could lead to the observed structure formation. The remarkable
agreement of the inflationary induced temperature fluctuations in the
Cosmic Microwave Background (CMB) radiation with all current
measurements is with no doubt an element which strongly supports
inflation.

Even though inflation provides a robust field theoretic mechanism
which answers the shortcomings of the hot big bang model and is in a
surprising agreement with CMB experiments, it still lacks a precise
theoretical model.  In this sense, inflation still remains a paradigm
in search of a model. One should keep in mind that an inflationary
model should be considered as successful when not only it fits the
data, but in addition it can be accommodated within (or even motivated
by) some fundamental theory (e.g., string theory).

In the context of supersymmetric grand unified theories, the most
natural inflationary model is that of hybrid inflation (Linde
1994). In this context, a detailed study (Jeannerot, Rocher \&
Sakellariadou 2003) of all Spontaneously Symmetry Breaking (SSB)
schemes which bring --- by one or more intermediate steps --- the
initial symmetric state of the universe, described by a large grand
unified theory gauge group, down to a less symmetric state, described
by the standard model gauge group, has shown that cosmic strings are
generically formed at the end of hybrid inflation (Sakellariadou
2008). Actually, in general such inflationary models require the
formation of cosmic strings so that inflation can indeed end (i.e.,
{\sl graceful exit}). Thus, cosmic strings have to be considered as a
{\sl sub-dominant partner} of inflation, and contribute (Bouchet
\textit{et al.} 2002) to the CMB temperature anisotropies. 

Current CMB data impose strong constraints in the maximum allowed
contribution of cosmic strings. More precisely, cosmic strings can
contribute at most 11$\%$ (Bouchet \textit{et al.}  2002; Wyman
\textit{et al.}  2005 ; Bevis \textit{et al.}  2008) to the power
spectrum of temperature anisotropies. This upper limit implies a
constrain in the string tension and therefore the energy scale of the
phase transition accompanied by SSB which left behind cosmic strings
as false vacuum remnants.  Equivalently, one can impose (Rocher \&
Sakellariadou 2005a,b, 2006) constraints on the parameter space
(couplings and mass scales) of the inflationary model (F-term, D-term
or P-term inflation). Cosmic strings formed in this way are sometimes
referred to as F-term or D-term strings, if produced at then end of F-
or D-term hybrid inflation, respectively\footnote{One should not {\sl
    identify} them with the Fundamental (F) strings or the
  one-dimensional Dirichlet branes (D-strings), respectively.}.

 Inflation must also prove itself generic, in the sense that the onset
 of inflation must be independent of initial conditions.  The first
 studies (Piran 1986; Goldwirth 1991; Calzetta \& Sakellariadou 1992,
 1993) addressing the onset of inflation were inconclusive, in the
 sense that no robust conclusions could be drawn as a quantum theory
 of gravity was missing. More recently, it has been shown (Germani,
 Nelson \& Sakellariadou 2007) that successful single-field inflation
 with a polynomial potential is highly improbable within the
 semi-classical regime of loop quantum cosmology\footnote{The
   probability to have an inflationary era with sufficient number of
   e-folds, is exponentially suppressed (Gibbons \& Turok 2006) in
   the context of standard general relativity.}. Inflation must
 be addressed in full quantum gravity, or in a string theory context.

It is at present widely believed that string theory is the fundamental
theory of all matter and forces, including a consistent quantum
gravity sector. If this is the case, then there must exist a natural
inflationary scenario within string theory. Such an approach will
allow the identification of the inflaton and the determination of its
properties, while at the same time cosmological measurements will
provide insight into the stringy description of our universe. Since
the {\sl discovery} of Dirichlet (D) branes, a natural realisation of
our universe in string theory is the brane-world scenario. In this
context, a simple, realistic and well-motivated inflationary model is
brane inflation, where inflation takes place while two branes move
towards each other, and their annihilation releases the brane tension
energy that heats up the universe to start the hot big bang
era. Typically, strings of all sizes and types may be produced during
the collision. Large Fundamental (F) strings and/or D1-branes
(D-strings) that survive the cosmological evolution become cosmic
superstrings. By observing strings in the sky, we will be able to
test, for the first (and maybe only) time, string theory.

Cosmic strings and superstrings --- thought for a long time completely
disconnected --- seem to be closely related, while inflation and
topological defects --- considered for many years either as
incompatible or as competing aspects of modern cosmology --- must
re-conciliate.

In what follows, I will highlight the most important properties of
cosmic superstrings.  I will first describe briefly the formation of
cosmic superstrings at the end of brane inflation. I will then discuss
the differences between cosmic superstrings and their solitonic
analogues, which arise in gauge theories. After a short review of the
evolution of Goto-Nambu cosmic strings, I will summarise the current
{\sl understanding} of the evolution of cosmic superstring
networks. This issue is far from being resolved.  The
characteristics of the superstring network are strongly dependent on
the specific brane inflationary model, while a realistic modelling
through numerical simulations is particularly complex.  Finally, I
will discuss the observational signatures of cosmic
strings/superstrings. This is a very promising area of research, since
it may give us the ({\sl unique}) way of testing string theory as a
realistic theory of nature.  There is a large number of open
directions which need to be thoroughly investigated and I will briefly
mention them. 

One should keep in mind that we are working in the context of type II
string theory.

\section{Cosmic superstring formation}
The possible astrophysical r\^ole of fundamental strings has been
advocated already more than twenty years ago. More precisely, it has
been proposed (Witten 1985), that superstrings of the O(32) and
E$_8\times$E$_8$ string theories are likely to generate string-like
stable vortex lines and flux tubes.  However, in the context of
perturbative string theory, the high tension (close to the Planck
scale) of fundamental strings ruled them out (Witten 1985) as
potential cosmic string candidates, for the following three reasons.
Firstly, such heavy strings would produce CMB inhomogeneities far
larger than the ones which have been measured.  Secondly, such high
tension strings could not have been produced after inflation, since
the scale of their tension is higher than the upper bound on the
energy scale of the inflationary vacuum. However, any topological
defects produced before inflation would have been diluted. Indeed, one
of the reasons for which inflation was suggested is to dilute unwanted
{\sl beasts} of the {\sl zoo} of topological defects; historically,
inflation was introduced to dilute monopoles. Thirdly, some
instabilities were identified, implying that such strings would not be
able to survive on cosmologically interested time-scales.

This picture has changed in the framework of brane-world cosmology,
which offers an elegant realisation of nature within string
theory. Within the brane-world picture, all standard model particles
are open string modes. Each end of an open string lies on a brane,
implying that all standard model particles are stuck on a stack of
D$p$-branes, while the remaining $p-3$ of the dimensions are wrapping
some cycles in the bulk. Closed string modes (e.g., dilaton, graviton)
live in the high-dimensional bulk.  Since gravity has been probed only
down to scales of about 0.1mm, the dimensions of the bulk can be much
larger than the string scale.  In the brane-world context, the
extra dimensions can even be infinite, if the geometry is non-trivial
and they are warped.  As a result of brane interactions, higher
dimensional D$p$-branes unwind and evaporate so that we are left with
D3-branes embedded in a (9+1)-dimensional bulk. One of these D3-branes
could play the r\^ole of our universe (Durrer, Kunz \& Sakellariadou
2005). Cosmic superstrings are also left behind.

Inflation can be easily accommodated in a string theory motivated
cosmological model.  String theory {\sl lives} in a high-dimensional
space, so compactification down to four space-time dimensions
introduces many gravitationally-coupled scalar fields --– moduli --– from
the point of view of the four-dimensional theory.  One of these fields
could play the r\^ole of the inflaton, provided it does not roll
quickly.  Runaway or light moduli are extremely problematic in
cosmology, so any realistic model should incorporate a mechanism for
moduli stabilisation. Indeed, a number of different approaches have
been followed (Giddings, Kachru \& Polchinski 2002; Kachru \textit{et
  al.}  2003; Kachru \textit{et al.} 2003) and various successful
inflationary models have been proposed within brane-world cosmology. 

A correct brane inflation scenario will offer us valuable information
on the early stages of our universe, as well as to the particular
compactification in string theory. String inflation models can be
classified according to the origin of the inflaton field. If the
inflaton is a scalar field arising from open strings ending of a
D$p$-brane ($p$ stands for the dimensionality of the Dirichlet brane),
then these open string models are called {\sl D-brane} (or just {\sl
  brane}) {\sl inflation models}. If the inflaton field is a moduli
(the most promising closed string modes), then these closed string
models are called {\sl moduli} or {\sl modular inflation}.

Brane annihilations can also provide a natural mechanism for ending
inflation. To illustrate the formation of cosmic superstrings at the
end of brane inflation, let us consider a D$p$-${\bar D}p$
brane-anti-brane pair annihilation to form a D$(p-2)$ brane.  Each
{\sl parent} brane has a U(1) gauge symmetry and the gauge group of
the pair is U(1)$\times$U(1). The {\sl daughter} brane possesses a
U(1) gauge group, which is a linear combination, U(1)$_-$, of the
original two U(1)'s. The branes move towards each other and as their
inter-brane separation decreases below a critical value, the tachyon
field, which is an open string mode stretched between the two branes,
develops an instability. The tachyon couples to the combination
U(1)$_-$. The rolling of the tachyon field leads to the decay of the
{\sl parent} branes. Tachyon rolling leads to spontaneously symmetry
breaking, which supports defects with even co-dimension.  So, brane
annihilation leads to vortices, D-strings; they are cosmologically
produced via the Kibble mechanism.  The other linear combination,
U(1)$_+$ disappears, since only one brane remains after the brane
collision.  The U(1)$_+$ combination is thought to disappear by having
its fluxes confined by fundamental closed strings.  Such strings are
of cosmological size and they could play the r\^ole of cosmic strings
(Sarangi \& Tye 2002; Jones, Stoica \& Tye 2003; Dvali \& Vilenkin
2004); they are referred to in the literature as cosmic superstrings
(Polchinski 2005).

Cosmic strings were found to be generically formed at the end of
inflation within supersymmetric grand unified theories. Likewise,
cosmic superstrings are found to be generically produced towards the
end of brane inflation.  The undesired, and cosmologically
catastrophic domain walls and monopole-like defects are not produced
within the type of string theory we are considering; in type IIB
string theory even dimensionality branes do not exist.

\section{Differences between cosmic strings and cosmic superstrings}

Solitonic cosmic strings are classical objects, which have been
traditionally assumed to share the characteristics of type-II
Abrikosov-Nielsen-Olesen (ANO) vortices (Abrikosov 1957; Nielsen \&
Olesen 1973) in the Abelian Higgs model.  Cosmic superstrings, despite
the fact that they are cosmologically extended, they are quantum
objects.  One thus expects differences in their properties, leading to
a different behaviour, and thus to distinct observational signatures.
 
Over distances that are large compared to the width of the string, but
small compared to the horizon size, solitonic cosmic strings can be
considered as one-dimensional objects and their motion can be
well-described by the Nambu-Goto action. However, this action cannot
be used to describe what happens when two strings intercommute; a
study which necessitates full field theory. When two strings of the
same type collide, they may either pass simply through one another, or
they may reconnect (intercommute). A necessary, but not sufficient,
condition for string reconnection is that the initial and final
configurations be kinematically allowed in the infinitely thing string
approximation.  Such a classical string solution for reconnection has
been shown to exist (Bettencourt \& Kibble 1994), but the precise
outcome of the string intersection depends on the internal structure
of strings.  Numerical simulations (and analytical estimates) of
type-II (and weakly type-I) strings in the Abelian Higgs model suggest
that the probability that a pair of strings will reconnect, after they
intersect, is close to unity (Shellard 1987; Laguna \& Matzner
1990). The results are based on lattice simulations of the
corresponding classical field configurations in the Abelian Higgs
model; the internal structure of strings is highly non-linear, and
thus difficult to treat via analytical means.  The only exception for which
string reconnection probability was found to be different than one is
the case of ANO strings with ultra-high collision speeds, in which
case they just pass through each other. In particular, for
near-perpendicular collisions the threshold speeds were found
(Ach\'ucarro \& de Putter 2006) to be bounded above by $\sim 0.97 c$
for type I and $\sim 0.90 c$ for type II strings.

The reconnection probability for cosmic superstrings is smaller (often
much smaller) than unity. The corresponding intercommutation
probabilities are calculated in string perturbation theory. The result
depends on the type of strings and on the details of
compactification. For fundamental strings, reconnection is a quantum
process and takes place with a probability of order $g_{\rm s}^2$
(where $g_{\rm s}$ denotes the string tension).  It can thus be much
less than one, leading to an increased density of strings
(Sakellariadou 2005), implying an enhancement of various observational
signatures.  The reconnection probability is a function of the
relative angle and velocity during the collision.  One may think that
strings can miss each other, as a result of their motion in the
compact space. Depending on the supersymmetric compactification,
strings can wander over the compact dimensions, thus missing each
other, effectively decreasing their reconnection probability.
However, it was found (Jackson, Jones \& Polchinski 2005) that in
realistic compactification schemes, strings are always confined by a
potential in the compact dimensions. The value of $g_{\rm s}$ and the
scale of the confining potential will determine the reconnection
probability. Even though these are not known, for a large number of
models it wass found (Jackson, Jones \& Polchinski 2005) that the
reconnection probability for F-F collisions lies in the range between
$10^{-3}$ and 1. The case of D-D collisions is more complicated; for
the same models the reconnection probability is anything between 0.1
to 1. Finally, the reconnection probability for F-D collisions can
vary from 0 to 1.

Brane collisions lead not only to the formation of F- and D-strings,
they also produce bound states, $(p,q)$-strings, which are composites
of $p$ F-strings and $q$ D-strings (Copeland, Meyers \& Polchinski
2004; Leblond and Tye 2004).  The presence of stable bound states
implies the existence of junctions, where two different types of
string meet at a point and form a bound state leading away from that
point.  Thus, when cosmic superstrings of different types collide,
they can not intercommute, instead they exchange partners and form a
junction at which three string segments. This is just a consequence of
charge conservation at the junction of colliding $(p, q)$-strings.
For $p = np'$ and $q = nq'$, the $(p, q)$ string is neutrally stable
to splitting into $n$ bound $(p', q')$ strings.  The angles at which
strings pointing into a vertex meet, is fixed by the requirement that
there be no force on the vertex. 

The formation of three-string junctions ($Y$-junctions) and kinematic
constraints for their collisions have been investigated analytically
(Copeland, Kibble \& Steer 2006, 2007; Copeland \textit{et al.} 2007),
under the assumption that each string evolves according to the
Nambu-Goto action. Whether these results hold for cosmic superstrings,
which carry fluxes of a gauge field and are therefore described
instead by the Dirac-Born-Infeld (DBI) action, remains to be shown.

The tension of solitonic strings is set from the energy scale of the
phase transition followed by a spontaneously broken symmetry which
left behind these defects as false vacuum remnants.  Cosmic
superstrings span a whole range of tensions, set from the particular
brane inflation model.  The tension of F-strings in 10 dimensions is
$\mu_{\rm F}=1/(2\pi\alpha')$, and the tension of D-strings is
  $\mu_{\rm D}=1/(2\pi\alpha' g_{\rm s})$, where $g_{\rm s}$ stands
  for the string coupling. In 10 flat dimensions, supersymmetry
  dictates that the tension of the $(p,q)$ bound states reads (Schwarz 1995)
\begin{equation}
\mu_{(p,q)}=\mu_{\rm F}\sqrt{p^2 + q^2/g_{\rm s}^2}~.
\label{squarelaw-tension}
\end{equation}
Individually, the F- and D-strings are ${1\over 2}$-BPS
(Bogomol'nyi-Prasad-Sommerfield) objects, which however break a
different half of the supersymmetry each. Equation
(\ref{squarelaw-tension}) represents the BPS bound for an object
carrying the charges of $p$ F-strings and $q$ D-strings.  In IIB
string theory, our universe can be described as a brane-world scenario
with flux compactification. In this context, the standard model
particles are light open string modes in a warped throat of the
Calabi-Yau manifold. The string tension for strings at the bottom of a
throat is different from the (simple) expression given in
Eq.~(\ref{squarelaw-tension}). The formula for tension depends on the
choice of flux compactification. For example, for the
Klebanov-Strassler throat (Klebanov \& Strassler 2000), inside which
the geometry is a shrinking $S^2$ fibered over a $S^3$, the tension
of the bound state of $p$ F-strings and that of $q$ D-strings reads
(Herzog \& Klebanov 2002; Hartnoll \& Portugues 2004; Gubser, Herzog
\& Klebanov 2004)
\begin{equation}
\mu_{\rm F_1}\simeq{h_{\rm A}^2\over 2\pi\alpha'}\ \sin\Big({\pi
  p\over M}\Big )\ \ \mbox{and}\ \ 
\mu_{\rm D_1}\simeq{h_{\rm A}^2\over 2\pi\alpha'}\ {q\over g_{\rm s}}~, 
\end{equation}
respectively, where $p$, $q$ are integers, $h_{\rm A}$ is the warp
factor at the bottom of the throat, $b=0.93$ is a number of the order
of unity, and $M$ denotes the number of fractional D3-branes.   The
tension formula for the $(p,q)$-bound states reads (Firouzjahi, Leblond \& Tye
2006)
\begin{equation}
\mu_{p,q}\simeq{h_{\rm A}^2\over 2\pi\alpha'}\sqrt{{q^2\over g_{\rm
      s}^2}+\Big({bM\over \pi}\Big)^2\sin^2\Big({\pi p\over M}\big)}~.
\label{bound-tension}
\end{equation}
For $M\rightarrow\infty$ and $b=h_{\rm A}=1$,
Eq.~(\ref{bound-tension}) reduces to Eq.~(\ref{squarelaw-tension}).

Type-I vortices in the Abelian Higgs model can also have three-vertex
junctions and a range of tensions (Donaire \& Rajantie 2006), in this
way they have more similarities with cosmic superstrings. However,
cosmic superstrings are the only ones to have the integer-valued
charges $p$ and $q$.  Thus, cosmic superstrings can, at least in
principle, be distinguished from gauge theory strings.

\section{Evolution of cosmic string/superstring networks}

Let me first summarise our understanding of the evolution of cosmic
string networks (Sakellariadou 2007). Knowing the differences between
cosmic strings and cosmic superstrings, it is at least in principle
{\sl easy} to identify possible deviations between the evolution of
the two networks.

The first studies of the evolution of a cosmic string network were
analytical. They have shown (Kibble 1985) the existence of {\sl
  scaling}, where at least the basic properties of the string network
can be characterised by a single length scale, roughly the persistence
length or the inter-string distance $\xi$ which grows with the
horizon.  This is a key property for cosmic strings since it renders
them cosmologically acceptable; a crucial difference between cosmic
strings and monopoles or domain walls.  The scaling solution was
supported by subsequent numerical work (Albrecht \& Turok 1985, 1989).
However, further investigation revealed dynamical processes, including
loop production, at scales much smaller than $\xi$ (Bennett \& Bouchet
1989; Sakellariadou \& Vilenkin 1990).  

The energy density of super-horizon ({\sl infinite}, or long) strings
in the scaling regime is given (in the radiation-dominated era) by
\begin{equation}
\rho_{\rm long} = \kappa\mu t^{-2} ~,
\end{equation}
where $\kappa$ is a numerical coefficient $(\kappa=20\pm 10)$.  The
sub-horizon loops, their size distribution, and the mechanism of their
formation remained for years the least understood parts of the string
evolution.  Assuming that the super-horizon strings are characterised
by a single length scale $\xi(t)$, one gets
\begin{equation}
\xi(t)=\Big({\rho_{\rm long}\over\mu}\Big)^{-1/2}=\kappa^{-1/2} t~.
\end{equation}
The typical distance between the nearest string segments and the
typical curvature radius of the strings are both of the order of
$\xi$.  

Early numerical simulations have shown that indeed the typical
curvature radius of long strings and the characteristic distance
between the strings are both comparable to the evolution time
$t$. Clearly, these results agree with the picture of the
scale-invariant evolution of the string network and with the one-scale
hypothesis.  However, the numerical simulations have also shown
(Bennett \& Bouchet 1988; Sakellariadou \& Vilenkin 1990) that
small-scale processes (such as the production of small sub-horizon
loops) play an essential r\^ole in the energy balance of long
strings. The existence of an important small-scale ({\sl wiggliness})
superimposed on the super-horizon strings was also evident by
analysing the string shapes (Sakellariadou \& Vilenkin 1990). 

In response to these findings, a three-scale model was developed
(Austin, Cope\-land \& Kibble 1993) which describes the network in
therms of three scales: the energy density scale $\xi$, a correlation
length $\bar\xi$ along the string, and a scale $\zeta$ relating to
local structure on the string. The small-scale structure (wiggliness),
which offers an explanation for the formation of the small sub-horizon
sized loops, is basically developed through intersections of long
string segments. It seemed likely from the three-scale model that
$\xi$ and $\bar\xi$ would scale, with $\zeta$ growing slowly, if at
all, until gravitational radiation became important when
$\zeta/\xi\approx 10^{-4}$ (Sakellariadou 1990; Hindmarsh 1990).
According to the three-scale model, the small length scale may reach
scaling only if one considers the gravitational back reaction
effect. Aspects of the three-scale model have been checked (Vincent,
Hindmarsh \& Sakellariadou 1997) evolving a cosmic string network is
Minkowski space-time. These string simulations found that loops are
produced with tiny sizes, which led the authors to suggest that the
dominant mode of energy loss of a cosmic string network is particle
production and not gravitational radiation as the loops collapse
almost immediately.

Recently, numerical simulations of cosmic string evolution in a
Friedmann-Lema\^{i}tre-Robertson-Walker universe, found evidence
(Ringeval, Sakellariadou \& Bouchet 2007) of a scaling regime for the
cosmic string loops in the radiation- and matter-dominated eras down
to the hundredth of the horizon time. It is important to note that the
scaling was found without considering any gravitational back reaction
effect; it was just the result of string intercmmutations. The scaling
regime of string loops appears after a transient relaxation era,
driven by a transient overproduction of string loops with length close
to the initial correlation length of the string network. Calculating
the amount of energy momentum tensor lost from the string network, it
was found (Ringeval, Sakellariadou \& Bouchet 2007) that a few
percents of the total string energy density disappear during a very
brief process of formation of numerically unresolved loops, which
takes place during the very first time-steps of the string
evolution. Subsequently, two other studies supported these findings
(Vanchurin, Olum \& Vilenkin 2006; Martins \& Shellard 2006).  More
recently, analytical studies (Polchinski \& Rocha 2006, 2007; Bubath,
Polchinski \& Rocha 2007) confirmed the numerical results of Ringeval
\textit{et al.} (Ringeval, Sakellariadou \& Bouchet 2007).

The evolution of cosmic superstrings is clearly a more involved
problem.  Cosmic superstring networks have not only sub-horizon loops
and super-horizon strings, they also have $Y$-junctions. This
complicates a lot the study of their evolution. In addition, one must
consider a multi-tension spectrum and reconnection probabilities which
can be much lower that unity. Certainly, computers are at present much
more efficient than in the eighties and nineties when we performed the
first numerical experiments with solitonic cosmic strings, and we
obviously gained a lot of experience from those studies. Nevertheless,
one must keep in mind that evolution of cosmic strings has been almost
exclusively studied in the (simple) case of the infinitely thin
approximation. So, even for the case of solitonic strings, the problem
is so complex that all numerical experiments have been performed for
the simplest (and probably, less realistic) models.

The evolution of a cosmic superstring network has very important
consequences for the validity of the brane inflation model
employed. The existence of $Y$-junctions may
prevent a scaling solution. If such a network freezes, it may
lead to undesirable cosmological consequences. A number of numerical
experiments have addressed (Sakellariadou 2005; Avgoustidis \&
Shellard 2005, 2006, 2007; Copeland \& Saffin 2005; Saffin 2005;
Hindmarsh \& Saffin 2006; Rajantie, Sakellariadou \& Stoica 2007;
Urrestilla \& Vilenkin 2007) this issue, each of them at a different
level of approximation.  I will briefly describe the approach and
findings of one of these numerical approaches (Rajantie, Sakellariadou
\& Stoica 2007), which I consider more realistic than others.  The aim
of that study was to build a simple field theory model of $(p,q)$
bound states, in analogy with the Abelian Higgs model used to
investigate the properties of solitonic cosmic string networks, and to
study the overall characteristics of the network using lattice
simulations.

The $(p,q)$ string network was modelled (Rajantie, Sakellariadou \&
Stoica 2007) using two sets of Abelian Higgs fields. Two models were
investigated, one in which both species of string have only
short-range interactions and another one in which one species of
string features long-range interactions.  More precisely, we modelled
the network with no long-range interactions using two sets of fields,
complex scalars coupled to gauge fields, with a potential chosen such
that the two types of strings will form bound states. This way
junctions of 3 strings with different tension was successfully
modelled.  In order to introduce long-range interactions we considered
a network in which one of the scalars forms global strings. This is
important if the strings are of a non-BPS species. For example, for
cosmic superstrings at the bottom of a Klebanov-Strassler throat the
F-string is not BPS while the D- string is. Thus, the different
components of the $(p,q)$ state are expected to exhibit different
types of long-range interactions.  The evolution of the string
networks suggested that the long-range interactions have a much more
important r\^ole in the network evolution than the formation of bound
states. In the local-global networks the bound states tend to split as
a result of the long-range interactions, resulting in two networks
that evolve almost independently. The formation of short-lived bound
states and their subsequent splitting only increases the small-scale
{\sl wiggliness} of the local strings. In the case of a local-local
network, the absence of long-range interactions allows the bound
states to be much longer-lived and significantly influences the
evolution of the string network (Rajantie, Sakellariadou \& Stoica
2007).

Even though preliminary studies indicate that the presence of
junctions is not itself inconsistent with scaling, this issue is far
from being resolved. Numerical experiments may support a scaling
solution, but this does not necessarily imply that realistic cosmic
superstring networks formed at the end of a successful and natural
brane inflationary era will reach scaling. Modelling a $(p,q)$ network
is indeed a challenging task, and further investigations are necessary
before this very important issue gets satisfactory answered.

\section{Observational consequences}

Since cosmic superstrings interact with the standard model particles
via gravity, their detection involves gravitational interactions. By
comparing the predictions of cosmic superstring models against recent
astrophysical data, one should in principle be able to constrain the
free parameters of the specific model. However, since the particular
brane inflationary scenario remains unknown, the tensions of
superstrings will be only loosely constrained.

I will briefly discuss the various observational signatures of cosmic
strings which have been studied in the (simple) case of the Abelian
Higgs model and relate them to those of cosmic superstrings. One
should keep in mind that all observational consequences of solitonic
strings have been studied in the (simple) case of the Abelian Higgs
model.  Even though curvature corrections have been investigated
(Gregory, Haws \& Garfinkle 1990; Letelier 1990; Barrab\`es, Boisseau,
\& Sakellariadou 1994), observational consequences of strings have been
only studied under the zero thickness assumption. The case of
superconduncting strings has been quite always neglected in the
studies of the observational signatures of cosmic strings. Under these
assumptions, cosmic strings can be well-described by the Goto-Nambu
action. In addition, the reconnection probability has been always
considered as exactly equal to unity. Finally, solitonic strings have
been always considered as having winding number equal to unity, thus
junctions are not formed.

Certainly, given the complexity of strings evolution (non-linear
process) it is very difficult to go beyond this simple context in
which solitonic strings have been studied. Nevertheless, one should be
very careful when one deduces the observational consequences of
strings; at least the quantitative discussion will be invalid for more
general string configurations.

The complexity of cosmic superstrings and the uncertainty of the
physical context (compactification, brane inflation model) of their
formation renders any discussion on their observational consequences
even more uncertain.

Cosmic strings perturb the space-time around them so that a conical
space-time is generated, leading to a unique gravitational lensing
signature through the appearance of undistorted double images.  The
finding of even a single such gravitational lensing event would be
seen as a convincing evidence for the existence of cosmic strings.  In
the case of strings with junctions, the $Y$-shaped junctions give rise
to lensing events that are qualitatively distinct from the case of
{\sl conventional} strings can produce (Shlear \& Wyman 2005;
Brandenberger, Firouzjahi \& Karouby 2007).  Idientifying such a
triple imaging event in the sky, would provide a {\sl smoking gun} for
the existence of a string network with non-trivial interactions.
Certainly, to generalise this study in the case of cosmic superstrings
may not be straightforward.

Micro-lensing is very useful in detecting lensing when the image
splitting is too small to resolve with astronomical measurements.
Gravitational micro-lensing of distant quasars by solitonic cosmic
strings has been recently investigated (Kuijken, Siemens \& Vachaspati
2007). The analysis seems to indicate rather pessimistic results for
the detectability of such micro-lensing events generated by cosmic
strings.  These studies have not been extended for strings
with non-trivial interactions.

Cosmic strings can also lead to weak gravitational lensing. A recent
study (Dyda \& Brandenberger 2007) on gauge cosmic strings has shown
that if such strings have a small-scale structure leading to a local
gravitational attractive force towards them, then an elliptical
distortion of the shape of background galaxies in the direction
corresponding to the projection of the string onto the sky may be
expected. Weak lensing has not been investigated in the case of cosmic
superstrings (neither in the simpler case of strings with
$Y$-junctions).

The CMB temperature anisotropies offer a powerful test for theoretical
models aiming at describing the early universe. The characteristics of
the CMB multipole moments can be used to discriminate among
theoretical models and to constrain the parameters space.  According
to our present understanding, the CMB temperature anisotropies
originate mainly from the amplification of quantum fluctuations at the
end of inflation, with a small contribution from the cosmic
(super)string network.  Given the small size of the observed CMB
temperature anisotropies, the perturbations may be treated linearly.
Thus, any coupling between perturbations induced by inflation and
those seeded by cosmic strings can be neglected.  Using the latest
WMAP data and Big Bang Nucleosynthesis (BBN) data, fractional
contribution from cosmic strings to the temperature power spectrum at
the multipole moment $\ell=10$ is at most 0.11 (Bevis \textit{et al.}
2008). In other words, if the normalisation of the string component
has been set to match the data at multipole $\ell=10$, the string
contribution cannot exceed 11\%.  This translates in the upper limit
on the dimensionless parameter $G\mu$ ($G$ is the gravitational
constant and $\mu$ the string tension) given by (Bevis \textit{et al.}
2008) $G\mu < 0.7\times 10^{-6}.$ This limit was derived for classical
Abelian Higgs strings with equal vector and scalar particle masses; it
is not expected to be valid for other types of strings.

The polarisation of the CMB photons can give further information and
constraints on the cosmological r\^ole of cosmic strings.  The B-mode
polarisation spectrum provides an important window on cosmic strings,
since the corresponding contribution from inflation is rather
weak. Scalar modes may contribute to the B-mode only via the
gravitational lensing of the E-mode signal, with a second inflationary
contribution arising from the sub-dominant tensor modes. It is thus
conceivable that the large vector contributions from cosmic strings
enable the detection of their imprint through future B-mode
measurements. In this way current views on {\sl natural} inflationary
models may challenge the conventional thought that a detection of
B-mode polarisation in the CMB will show the existence of gravity
waves in the early universe and determine the energy scale of
inflation. To distinguish the cosmic string from the inflationary
gravity wave signal one should go to rather high energy resolution,
since the signal from cosmic strings seem to be dominant at $\ell\sim
1000$, while the gravity wave signal from inflation peaks at $\ell\sim
100$ (Seljak \& Slosar 2006).  The prediction of a large cosmic string
contribution to the B-mode polarisation power spectrum has been
confirmed even for small string contributions to the CMB.  More
precisely, it has been argued (Bevis \textit{et al.} 2007) that data
from future ground-based polarisation detectors may bound the
dimensionless string parameter to $G\mu <0.12\times 10^{-6}$.

Cosmic strings can also become apparent through their contribution in
the small-angle CMB temperature anisotropies.  More precisely, at high
multipoles $\ell$ (small angular resolution), the mean angular power
spectrum of string-induced CMB temperature anisotropies can be
described (Fraisse \textit{et al.} 2007) by $\ell^{-\alpha}$, with
$\alpha \sim 0.889$.  Thus, a non-vanishing string contribution to the
overall CMB temperature anisotropies may dominate at high multipoles
$\ell$ (small angular scales).  In an arc-minute resolution
experiment, strings may be observable for $G\mu$ down to $2\times
10^{-7}$ (Fraisse \textit{et al.} 2007). 

Cosmic strings should also induce deviations from Gaussianity. On
large angular scales such deviations are washed out due to the low
string contribution, however on small angular scales, optimal
non-Gaussian, string-devoted statistical estimators may impose severe
constraints on a possible cosmic string contribution to the CMB
temperature anisotropies.

One should keep in mind that all string-induced CMB temperature
anisotropies were performed for Abelian strings in the zero thickness
limit with reconnection probability equal to unity and winding number
equal to one. Even though in any model where fluctuations are
constantly induced by sources ({\sl seeds}), having a non-linear
evolution, the perfect coherence which characterises the inflationary
induced spectrum of perturbations gets destroyed (Durrer \textit{et
  al.} 1997; Durrer \& Sakellariadou 1997), there is still no reason to
expect that quantitatively the results found for conventional cosmic
strings models will hold in more general cases.

Cosmic strings are expected to produce a stochastic background of
Gravity Waves (GW), which can be estimated by the incoherent
superposition of GW bursts at the cusps and kinks of a network of
oscillating string loops. This stochastic GW background, may be
detectable by pulsar timing observations.  Gravity waves bursts
emitted from cusps of oscillating string loops, may be detected by the
LIGO/VIRGO and LISA interferometers.  For {\sl conventional} cosmic
strings it has been argued (Damour \& Vilenkin 2000, 2001) that even
if only 10\% of all string loops have cusps, the GW bursts might be
detectable by the planned GW detectors LIGO/VIRGO and LISA for string
tensions as small as $G\mu \sim∼ 10^{-13}$. The result depends on the
number of cusps which is still not well-known.

Some preliminary studies of gravity wave emission from cosmic
superstring networks have been recently performed (Damour \& Vilenkin
2005). The only difference which has been considered for those
investigations is the reconnection probability. These studies seem to
indicate (Damour \& Vilenkin 2005) that the smaller reconnection
probability will enhance the observational signature of cosmic
superstrings. Analysing the BBN, CMB and pulsar timing bounds, it
seems that the BBN and CMB bounds are consistent with, but somewhat
weaker than, the pulsar bound.  It is argued (Siemens \textit{et al.}
2006a,b) that considering string networks with small reconnection
probabilities, ${\cal P}$, strings with $G\mu \geq 10^{-12}$ are ruled
out when ${\cal P} \sim 10^{-3}$. Increasing the reconnection
probability, strings with $G\mu\geq 10^{-10}$ are ruled out, while the
bound becomes $G\mu\geq 10^{-8}$ for ${\cal P}\sim 1$. These results
depend on the evolution of the string network, the number of
cusps/kinks, as well as the reconnection probability. Since only the
reconnection probability has been taken into account, it is rather
immature to claim that these bounds correspond to realistic cosmic
superstring networks. Further investigations are necessary.

The energy scale of cosmic strings/superstrings can be also
constrained from the emission of moduli. Preliminary studies have been
done (Damour \& Vilenkin 1997; Sakellariadou 2005; Babichev, \&
Kachelriess 2005; Davis, Binetruy \& Davis 2005; Firouzjahi 2007),
however further studies with more realistic cosmic superstring
networks are required.

\section{Conclusions}

For a number of years, inflation and topological defects have been
considered either as two incompatible or as two competing aspects of
modern cosmology. Historically, one of the reasons for which inflation
was proposed is to rescue the standard hot big bang model from the
monopole problem. However, such a mechanism could also dilute cosmic
strings unless they were produced at the end or after inflation. Later
on, inflation and topological defects competed as the two alternative
mechanisms generating the density perturbations which led to the
observed large-scale structure and the anisotropies in the CMB.  The
plethora of data on the CMB have revealed an early universe in
striking agreement with the basic predictions of inflation, while
there is a clear inconsistency between predictions from topological
defect models and CMB data. This indicates a clear preference for
inflation. However, despite its success inflation still lacks a
precise theoretical model.

In the context of a brane-world cosmological scenario within string
theory, brane inflation can be easily accommodated.  Brane inflation
takes place while two branes move towards each other, and their
annihilation releases the brane tension energy that heats up the
universe to start the radiation-dominated era of the hot big bang
model. Typically, strings of all sizes and types may be produced
during the collision.  Large fundamental strings and/or D1-branes that
survive the cosmological evolution become cosmic superstrings.

The study of cosmic superstrings and their observational signatures
will give us an understanding of the early stages of our universe and
it will provide information in identifying the details of the string
theory model which has relevance with our universe. Thus, the issue of
cosmic suprestrings is gaining a lot of interest from the scientific
community.  Numerical experiments of cosmic superstring networks and
analytical studies are trying to unravel the properties and
characteristics of such networks.  The problem is quite complex and
our intuition from the (better known) {\sl conventional} solitonic
strings described by the Goto-Nambu action, may turn out to be
misleading. Many open questions are currently under investigation and
we expect to have soon a better understanding of their properties and
their observational signatures which will allow us to detect them.

\begin{acknowledgements}
It is a pleasure to thank the organisers of the Royal Society
Discussion Meeting, {\sl Cosmology meets condensed matter}, who
invited me to give this lecture. This work was supported in part by
the European Union through the Marie Curie Research and Training
Network UniverseNet (MRTN-CN- 2006-035863).

\end{acknowledgements}

\label{lastpage}
\end{document}